\documentclass[12pt,preprint]{aastex}

\shorttitle{Nonthermal Emission from Cluster Centers}
\shortauthors{Fujita et al.}

\begin{document}

\title{Nonthermal Emission Associated with Strong AGN Outbursts
at the Centers of Galaxy Clusters}

\author{Yutaka Fujita\altaffilmark{1}, Kazunori Kohri\altaffilmark{2,3}, 
Ryo Yamazaki\altaffilmark{4},
and Motoki Kino\altaffilmark{1,5}}

\altaffiltext{1}{Department of Earth and Space Science, Graduate School
of Science, Osaka University, 1-1 Machikaneyama-cho, Toyonaka, Osaka
560-0043, Japan; fujita@vega.ess.sci.osaka-u.ac.jp} 
\altaffiltext{2}{Institute for Theory and Computation,
Harvard-Smithsonian Center for Astrophysics, MS-51, 60 Garden Street,
Cambridge, MA 02138} 
\altaffiltext{3}{Physics Department, Lancaster
University, Lancaster LA1 4YB, UK; k.kohri@lancaster.ac.uk} 
\altaffiltext{4}{Department of
Physics, Hiroshima University, Higashi-Hiroshima, Hiroshima 739-8526,
Japan; ryo@theo.phys.sci.hiroshima-u.ac.jp}
\altaffiltext{5}{Institute of Space and Astronautical Science, JAXA, 
3-1-1 Yoshinodai, Sagamihara, Kanagawa 229-8510, Japan; 
kino@vsop.isas.jaxa.jp}

\begin{abstract}
 Recently, strong AGN outbursts at the centers of galaxy clusters have
 been found. Using a simple model, we study particle acceleration around
 a shock excited by an outburst and estimate nonthermal emission from
 the accelerated particles. We show that emission from secondary
 electrons is consistent with the radio observations of the minihalo in
 the Perseus cluster, if there was a strong AGN outburst $\gtrsim
 10^8$~yrs ago with an energy of $\sim 1.8\times 10^{62}$~erg. The
 validity of our model depends on the frequency of the large
 outbursts. We also estimate gamma-ray emission from the accelerated
 particles and show that it could be detected with {\it GLAST}.
\end{abstract}


\section{Introduction}

Diffuse synchrotron radio emission is often found in the intracluster
medium (ICM) of galaxy clusters
\citep[e.g.,][]{kim90,gio93,gio00,kem01}. These radio sources in
clusters are often classified as either peripheral cluster radio relic
sources or central cluster radio halo sources. Since most of the relics
and halos are found in merging clusters, the particles responsible for
the emission are thought to be accelerated at shocks or turbulence
excited by cluster mergers
\citep*{roe99,tak00,fuj01,bru01,ohn02,fuj03,bru04}.

However, there are exceptional diffuse radio sources. They are called
`minihalos' and are located in the central regions of non-merging
clusters or `cooling core' clusters
\citep{bau91,bur92,riz00}. \citet{git02} suggested that the diffuse
synchrotron emission from radio minihalos is due to a relic population
of relativistic electrons reaccelerated by MHD turbulence via
second-order Fermi acceleration, and that the energy is supplied by
cooling flows. Alternatively, \citet{pfr04} discussed that the
electrons responsible for the synchrotron emission from minihalos are of
secondary origin and thus are injected during proton-proton collision in
the ICM. However, \citet{git02} did not discuss the generation of the
turbulence, and \citet{pfr04} did not investigate the mechanism of
proton acceleration.

Recent X-ray observations have shown that AGNs at the centers of galaxy
clusters sometimes exhibit intensive outbursts with a mechanical power
of $\sim 10^{61}$~erg. The examples are MS~0735.6+7421
\citep{mcn05,git07}, Hercules~A \citep{nul05a}, and Hydra-A
\citep{nul05b,wis07}. Such an intensive outburst should excite a shock
in the ICM. In fact, weak shocks have been found in those clusters. In
the early stage of the evolution of the shock, the Mach number is
expected to be large.  Therefore, particles would be accelerated at the
shock as in the case of a supernova remnant.

In this letter, we consider particle acceleration at the shock generated
by an intensive outburst of the AGN at the cluster center. We study
nonthermal emission from the accelerated particles. In particular, we
focus on nonthermal radio emission of secondary origin. Recently,
\citet{hin07} estimated gamma-ray emission associated with AGN
outbursts, assuming that the hot cavity behind a shock is entirely
filled with high-energy protons. This assumption may be too simple, and
they did not discuss the acceleration and energy spectrum of
protons. However, motivated by this study, we also estimate the
gamma-ray emission using our model. We take the Perseus cluster as a
model cluster, because this cluster has a well-studied minihalo.

\section{Models}

We assume that the duration of an AGN outburst is $\sim 10^7$~yr. Since
we are interested in the evolution of a shock for $t\gtrsim 10^7$~yr,
where $t=0$ corresponds to the ignition of the outburst, the evolution
can be approximated by that of an instant explosion with an energy of
$E_0$ at $t=0$.

For the sake of simplicity, we assume that the cluster is spherically
symmetric and the density profile of the ICM has a form of a power-law:
\begin{equation}
 \rho_{\rm ICM}(r)=\rho_1(r/r_1)^{-\omega}\:.
\end{equation}
We take $\rho_1=5.3\times 10^{-25}\rm\: g\: cm^{-3}$, $r_1=10$~kpc, and
$\omega=1.43$, based on the density profile of the Perseus cluster for
$70< r<300$~kpc \citep[Fig.~8 in][]{chu03}. We take that region because
our model is correct for $t\gtrsim 1\times 10^7$~yr, and the radius of
the shock at $t\sim 10^7$~yr is $R_s\sim 70$~kpc for parameters we adopt
in \S\ref{sec:results}.

Using a shell approximation \citep[e.g.][]{ost88}, the radius of the
shock can be written as
\begin{equation}
\label{eq:Rs}
 R_s = \xi\left(\frac{E_0}{\rho_1 r_1^\omega}\right)^{1/(5-\omega)}
t^{2/(5-\omega)}\;,
\end{equation}
where
\begin{equation}
 \xi = \left[\left(\frac{5-\omega}{2}\right)^2\frac{3}{4\pi}
\frac{(\gamma+1)^2(\gamma-1)(3-\omega)}{9\gamma-3-\omega(\gamma+1)}
\right]^{1/(5-\omega)}\:,
\end{equation}
and $\gamma (=5/3)$ is the adiabatic index.  The Mach number of the
shock gradually decreases. The shock stops expanding when its Mach
number approaches to one. At this point, the cavity filled with hot gas
inside the shock becomes in pressure equilibrium with the surrounding
ICM. Since we stop calculation before the radiative cooling of the shock
becomes effective ($\gtrsim$~Gyr), we do not need to consider the
radiative cooling.

Following \citet{yam06}, we assume that particles are accelerated at the
shock via diffusive shock acceleration (i.e., first-order Fermi
acceleration) and that their energy spectra are given by
\begin{equation}
 N(E)\propto E^{-x}e^{-E/E_{\rm max}}\:,
\label{eq:NE}
\end{equation}
where $E_{\rm max}$ is the maximum energy of the protons or electrons.
The index is given by $x=(r_b+2)/(r_b-1)$, where $r_b$ is the
compression ratio of the shock \citep{bla87}. We estimate the maximum
energies of the protons and electrons using the relations of
\begin{equation}
 t_{\rm acc}=\min\{t_{pp},t\}\:, \hspace{10mm} 
t_{\rm acc}=\min\{t_{\rm syn},t\}\:,
\label{eq:rel}
\end{equation}
respectively. Here, $t_{\rm acc}$, $t_{pp}$, $t$ and $t_{\rm syn}$ are
the acceleration time, the lifetime of high-energy protons through pion
production, the age of the shock wave, and the synchrotron cooling time,
respectively.

Assuming the standard manner for the diffusion coefficient, the
acceleration time is given by
\begin{equation}
t_{acc}=\frac{20hc E_{\rm max}}{e B_d V_s^2}\:,
\end{equation}
where $c$ is the speed of light, $-e$ is the electron charge, and $V_s
(=dR_s/dt)$ is the shock velocity \citep{jok87,yam04}. The correction
factor $h$ depends on the mean free path of particles and the angle
between the shock and the magnetic field. Since $h\sim 1$ in the
Bohm-limit case, we assume that $h=1$ from now on. The downstream
magnetic field is given by $B_d = r_b B$, where $B$ is the magnetic
field strength of the unperturbed ICM.  We estimate $t_{pp}$ as
\begin{equation}
 t_{pp}=5.3\times 10^7 \:{\rm yr}\:
(n_{\rm ICM}/\rm cm^{-3})^{-1},
\label{eq:tpp}
\end{equation}
where $n_{\rm ICM}$ is the number density of the ICM.  Since the shock
is in pressure equilibrium in $\sim10^8$~yr (see \S\ref{sec:results})
and $n_{\rm ICM}\lesssim 0.1\rm\: cm^{-3}$, the cooling is not
effective. Thus, the maximum energy of protons is determined by the age
of the shock. On the other hand, the synchrotron cooling time for
electrons is given by
\begin{equation}
 t_{\rm syn}=1.25\times 10^4\:{\rm yr}\:\left(\frac{E_{\rm max,e}}{10\rm\:
			   TeV}\right)^{-1} \left(\frac{B_d}{10\mu\:\rm
			   G}\right)^{-2},
\end{equation}
and is shorter than the age of the shock.
Thus, using relations~(\ref{eq:rel}), we obtain
\begin{equation}
 E_{\rm max, p}\sim 1.6\times 10^2\: 
\left(\frac{V_s}{10^3\rm\: km\: s^{-1}}\right)^2
\left(\frac{B_d}{10\: \mu\rm\: G}\right)
\left(\frac{t}{10^5\:\rm yr}\right)\:{\rm TeV}\:,
\label{eq:pmax}
\end{equation}
\begin{equation}
 E_{\rm max, e}\sim 14\:
\left(\frac{V_s}{10^3\rm\: km\: s^{-1}}\right)
\left(\frac{B_d}{10\: \mu\rm\: G}\right)^{-1/2}\:{\rm TeV}\:.
\label{eq:emax}
\end{equation}

We assume that the minimum electron and proton energies are their rest
masses. For given proton and electron spectra, we calculate radiation
from them. We consider the synchrotron, bremsstrahlung, and inverse
Compton emissions from primary electrons that are directly accelerated
at the shock, the $\pi^0$-decay gamma-ray through proton-proton
collisions, and the synchrotron, bremsstrahlung, and inverse Compton
emissions from secondary electrons created through the decay of charged
pions that are also generated through proton-proton collisions
\citep*{stu97,koh07}. The density of target protons for the
proton-proton interaction is given by $r_b\rho_{\rm ICM}(R_s)/(1.4\:
m_p)$, where $m_p$ is the proton mass. We assume that the spectrum from
secondary electrons is stationary if the lifetime of the electrons is
smaller than the age of the system. On the other hand, if the lifetime
is larger than the system age, we calculate the evolution according to
\S3 of \citet{ato99} \citep[see also][]{koh07}.

\section{Results}
\label{sec:results}

In our model, the evolution of a shock is determined by $\rho_{\rm ICM}$
and $E_0$ (equation~[\ref{eq:Rs}]). The Mach number also depends on the
ICM temperature, $T$. The energy spectrum of particles depends on the
evolution of the shock and the magnetic field, $B$. The luminosity of
nonthermal emission from the shock depends on the total energy of
high-energy ($>m_p c^2$) protons inside the shock, $\epsilon E_0$, where
$0\leq\epsilon \leq 1$. We fix $\rho_{\rm ICM}(r)$, $T$, and $B(r)$ from
observations. On the other hand, we regard $E_0$ and $\epsilon$ as
fitting parameters, because there are no observational data for them.

We assume that $T=3.5$~keV, which is the temperature of the central
region of the Perseus cluster \citep{chu03}, although the temperature
before the outburst might have been somewhat lower. As far as we know,
there are no observations of magnetic fields in the Perseus cluster at
$r\gtrsim 70$~kpc. On the other hand, the observations of Faraday
rotation showed that the typical magnetic field strength in clusters for
$r\lesssim 500$~kpc is 5--$10\: \mu$G \citep*{cla01a}. Therefore, we
take $B(r)=7\mu{\rm G}\:(\rho_{\rm ICM}[r]/\rho_{\rm ICM}[150 \:\rm
kpc])^{2/3}$ assuming that the magnetic field is adiabatically
compressed. We note that the spectra of particles
(equations~[\ref{eq:pmax}] and~[\ref{eq:emax}]) and synchrotron emission
from high-energy electrons depend on $B$. Thus, the results shown below
is fairly sensitive to the assumption on $B$.

In the following, the energy of an AGN explosion is $E_0=1.8\times
10^{62}\rm\: erg$, which is three times larger than the one observed for
MS~0735.6+7421 \citep{mcn05}. We use this value to match $R_s$ with the
size of the radio minihalo in the Perseus cluster. We take the
acceleration efficiency of $\epsilon=0.05$ to match radio observations
(see below). The ratio of high-energy electrons to high-energy protons
is taken to be $r_{e-p}=1/1000$ as a fiducial value. In the self-similar
solution of the shock we adopt, the kinetic and thermal energies are
respectively constant. Therefore, we assume that the total energy of the
high-energy protons is also constant. The typical Mach number of the
shock in our calculations is $\sim 3$ (for $1\times 10^7<t<4\times
10^7$~yr). Performing simulations taking account of the back-reaction of
accelerated particles on hydrodynamics, \citet{ryu03} estimated that the
cosmic-ray acceleration efficiency is $\sim 0.2$ for that Mach
number. Since they defined cosmic-ray as particles with energies larger
than those of thermal particles (or the injection energy for
acceleration), the fraction of protons having energies of $>m_p c^2$
must be smaller than 0.2. Although it is not certain whether
equation~(\ref{eq:NE}) can be extrapolated down to the injection energy,
the adopted value of $\epsilon=0.05$ is consistent with that of
\citet{ryu03} because it is smaller than 0.2.

Fig.~\ref{fig:0.02} shows the spectrum of nonthermal emission from
accelerated particles at $t=2\times 10^7$~yr. The distance to the model
cluster is 78.4~Mpc (the distance to the Perseus cluster\footnote{The
redshift of the Perseus cluster is 0.0183. We assumed that the
cosmological parameters are $\Omega_0=0.3$, $\lambda_0=0.7$, and
$H_0=70\:\rm km\: s^{-1}\: Mpc^{-1}$}).  Synchrotron emission from
primary electrons is dominant upto $\sim 100$~keV. The maximum energies
for protons and electrons are $E_{\rm max,p}=6.8\times 10^{17}$~eV and
$E_{\rm max, e}=2.1\times 10^{13}$~eV, respectively. The radius of the
shock at this time is $R_s=97$~kpc, the shock velocity is $V_s=2650\rm\:
km\: s^{-1}$, and the Mach number is 2.7.

Fig.~\ref{fig:0.04} shows the spectrum at $t=4\times 10^7$~yr. At this
time, $E_{\rm max,p}=4.1\times 10^{17}$~eV, $E_{\rm max, e}=2.1\times
10^{13}$~eV, and $R_s=143$~kpc, which is close to the size of the
minihalo in the Perseus cluster. The shock velocity is $V_s=1950\rm\:
km\: s^{-1}$ and the Mach number is 2.0. In Figs.~\ref{fig:0.02} and
\ref{fig:0.04}, we plot the observed radio fluxes of the minihalo in the
Perseus cluster \citep{sij93,git02}. As can be seen, the predicted radio
synchrotron emission (long-dashed line) is too bright to be consistent
with the observations.  If we take smaller $E_0$, the radio luminosity
becomes smaller. However, the size of the minihalo is too small to be
consistent with the observed one.  Moreover, if we consider a much
smaller electron-proton ratio (e.g. $r_{e-p}\sim 10^{-5}$), the radio
spectral index is inconsistent with the observations.

One possibility is that the age is much larger than $4\times 10^7$~yr
and is $t\gtrsim 10^8$~yr. At that time, the shock is not prominent
because it is almost in pressure equilibrium with the surrounding ICM.
In fact, for the Perseus cluster, a shock of $R_s\sim 100$--200~kpc has
not been reported.  As the shock expands, its Mach number decreases from
2.7 (at $t=2\times 10^7$~yr) to 2.0 (at $t=4\times
10^7$~yr). Accordingly, the compression ratio ($r_b$) decreases, which
affects the energy spectrum of particles and the emission from them
(eq.~[\ref{eq:NE}]). \citet{ryu03} indicated that particles are no
longer accelerated if the Mach number is $\lesssim 2$. Thus, for
$t\gtrsim 4\times 10^7$~yr, particle acceleration at the shock is not
effective. 

At $t\sim 10^8$~yr, the emission from primary electrons may have died
out, because electrons with a Lorentz factor of $\gamma\gtrsim 10^4$,
which are responsible for the radio emission, lose their energy through
synchrotron emission and inverse Compton emission on a time-scale of
$\lesssim 4\times 10^7$ \citep[e.g.][]{sar99}. On the other hand, the
lifetime of protons is much larger than $10^8$~yr (eq.~[\ref{eq:tpp}]),
and the diffusion time of protons having the maximum energy of $\sim
10^{17}$~eV from the central region of the cluster ($\sim 200$~kpc) is
$2\times 10^8$~yr \citep*{vol96}. Since particles are no longer
accelerated at $t\gtrsim 4\times 10^7$~yr, the overall spectrum
originated from proton-proton collisions stays almost intact for
$4\times 10^7\lesssim t \lesssim 2\times 10^8$~yr. In the radio band,
only synchrotron emission from secondary electrons (dot-dashed line in
Fig.~\ref{fig:0.04}) will be observed at $t\sim 10^8$~yr. In this case,
the predicted spectrum (thick-solid line in Fig.~\ref{fig:0.04}) well
fits the radio observations. Since the secondary radio emission is
produced by protons having energies of $\sim 100$~GeV and since the
diffusion time of these protons is $>$~Gyr, the radio emission could
persist that time. We emphasize that the assumption on the
electron-proton ratio ($r_{e-p}$) is not required to estimate the
emission originated from proton collisions.

\section{Discussion}
\label{sec:dis}

\begin{figure}
\plotone{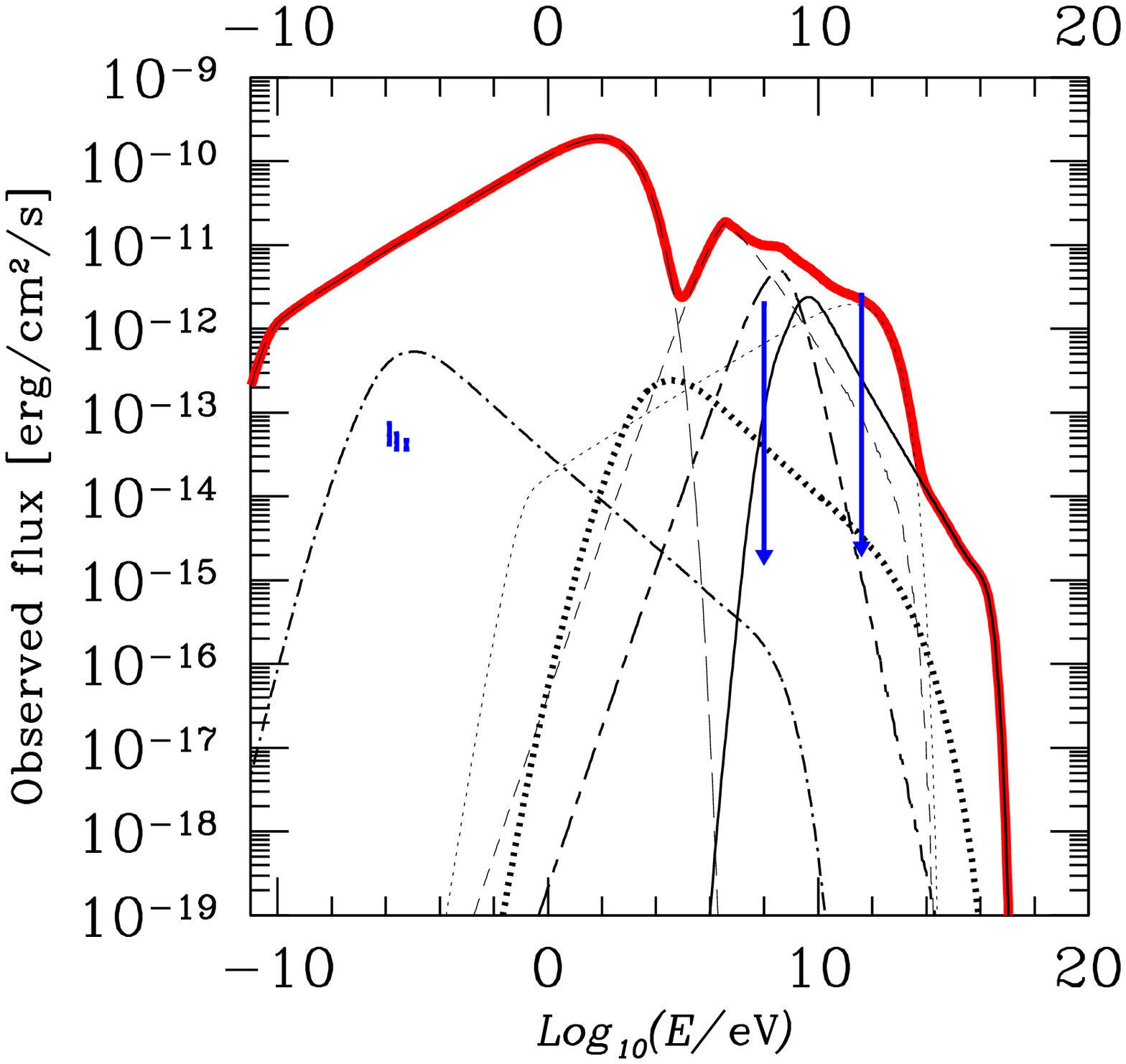} \caption{The spectrum of a shock at $t=2\times
 10^7$~yr. Emissions from primary electrons are synchrotron
 (long-dashed), bremsstrahlung (short-dashed) and inverse Compton (thin
 dotted). Emissions related to protons are $\pi^0$-decay gamma-ray
 (thin-solid), synchrotron (dot-dashed), bremsstrahlung (short-and-long
 dashed), and inverse Compton (thick-dotted) emissions from secondary
 electrons. The thick-solid line shows the total nonthermal flux. Radio
 observations are shown by dots \citep{sij93,git02}, and gamma-ray upper
 limits are shown by arrows \citep{per06}. \label{fig:0.02}}
\end{figure}

\begin{figure}
\plotone{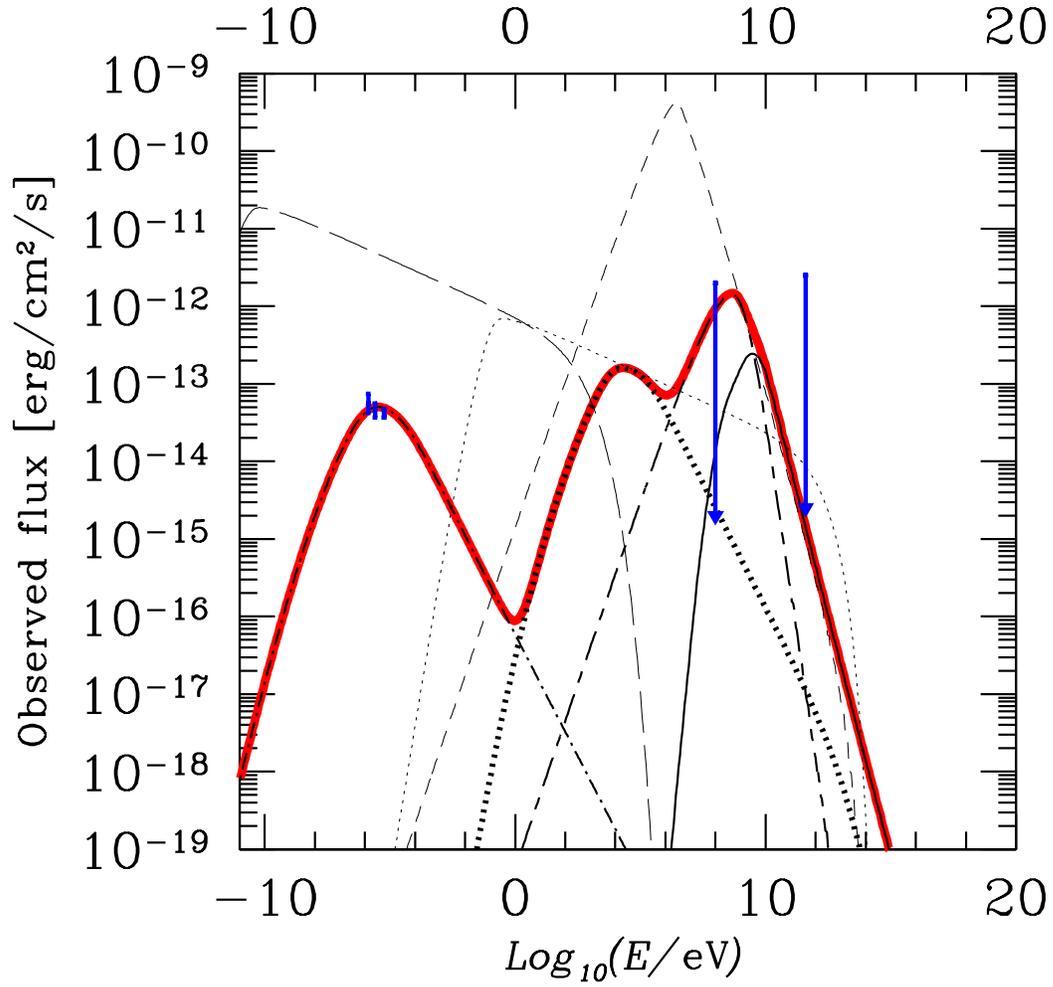} \caption{Same as Fig.~\ref{fig:0.02}, but for
 $t=4\times 10^7$~yr. In contrast with Fig.~\ref{fig:0.02}, the
 thick-solid line shows the flux from protons.\label{fig:0.04}}
\end{figure}

Although our model is basically an one-zone model and cannot
quantitatively predict the spatial change of the spectrum, we can
qualitatively predict that. Compared with the one at $t=2\times 10^7$~yr
(Fig.~\ref{fig:0.02}), the spectrum of synchrotron emission from
secondary electrons at $t=4\times 10^7$~yr is softer in the radio band
($\sim 0.3$--1~GHz; Fig.~\ref{fig:0.04}). The spectral index in the band
of 327--609~MHz changes from 1.64 ($t=2\times 10^7$~yr, $R_s=97$~kpc) to
1.88 ($t=4\times 10^7$~yr, $R_s=142$~kpc). Since some of the protons
accelerated at an earlier time should remain in the inner region of a
cluster, the spectrum should be less steep in the inner region. This
tendency is consistent with observations \citep{sij93,git02}. 

As we mentioned above, the radio emission from secondary electrons could
persist for a long time ($>$~Gyr). Our model will be tested for the
frequency (or the event rate) of large outbursts. \citet{git07}
indicated that large outbursts are likely occurring $\sim 10$\% of the
time in a significant proportion of all cooling core clusters. Thus, our
model would suggest that a large fraction of clusters should have
minihalos. This is inconsistent with the rareness of minihalos. However,
the outbursts observed so far are of energies of $<10^{62}\rm\: erg$
\citep{raf06}, which is smaller than our finding ($1.8\times
10^{62}$~erg). Thus, the rareness may indicate that strong AGN outbursts
with energies of $> 10^{62}\rm\: erg$ are rare phenomena or minor
cluster mergers often perturb cluster cores. Another possibility is that
particle acceleration at low-Mach number shocks occurs only in some
specific IGM environments depending on the density of the surrounding
matter, magnetic field configurations, and so on. In the future,
statistical studies about AGN outbursts of $\sim 10^{62}\rm\: erg$ are
highly desired. The morphology of the radio surface brightness would
also be important to check the validity of the model. If the particle
acceleration is triggered by the expanding shock, one would expect a
torus-like shape instead of the spherical shape observed for minihalos
\citep{git02}. However, for clusters observed so far, the central region
behind the shock is not empty; the ICM is still filling
\citep[e.g. Fig.~3 of][]{mcn05}. Thermal protons there may work as
target protons for the proton-proton interaction and thus the radio
emission may not be a torus-like shape. A spatially resolved model must
be constructed to address this issue.

In Figs.~\ref{fig:0.02} and \ref{fig:0.04}, we also plot the
observational upper limits of gamma-ray emission from the Perseus
cluster \citep{per06}. At $t=2\times 10^7$~yr, the gamma-ray emission is
brighter than the observations. At $t\sim 10^8$~yr, there is no longer
emission from primary electrons, and only gamma-ray emission of proton
origin (thick-solid line in Fig.~\ref{fig:0.04}) will be observed in the
gamma-ray band. The predicted gamma-ray flux at $E\sim 10^9$~eV is $\sim
1\times 10^{-12}\:\rm erg\: cm^{-2}\: s^{-1}$, which could be detected
with {\it
GLAST}~\footnote{http://www-glast.slac.stanford.edu/software/IS/glast\%5Flat\%5Fperformance.htm}
with a sensitivity of $\sim 3\times 10^{-13}\:\rm erg\: cm^{-2}\:
s^{-1}$. The gamma-ray emission would persist for $\sim t_{pp}$. If the
gamma-ray is detected, it directly indicates that protons as well as
electrons are accelerated in clusters. Moreover, the luminosity reflects
the total energy of the protons.

On the other hand, it would be difficult to detect the emission with
imaging atmospheric Cherenkov telescopes. For example, {\it
H.E.S.S.}\footnote{http://www.mpi-hd.mpg.de/hfm/HESS/HESS.html} has a
sensitivity of $\sim 1\times 10^{-13}\:\rm erg\: cm^{-2}\: s^{-1}$ at
$\sim 10^{12}$~eV). The predicted flux is smaller than the detection
limit (Fig.~\ref{fig:0.04}).

\acknowledgments

The authors wish to thank the referee for useful comments. We are also
grateful to T. Mizuno and Y. Ohira for fruitful discussions. Y.~F.\ and
R.~Y.\ were each supported in part by Grants-in-Aid from the Ministry of
Education, Science, Sports, and Culture of Japan (Y.~F.: 17740162,
R.~Y.: 18740153). K.~K. was also supported in part by NASA grant
NNG04GL38G, PPARC grant, PP/D000394/1, EU grant MRTN-CT-2006-035863, the
European Union through the Marie Curie Research and Training Network
"UniverseNet" (MRTN-CT-2006-035863)

\clearpage

\clearpage

\end{document}